\documentclass[aps,prmaterials,twocolumn,groupedaddress,showpacs,notitlepage]{revtex4-1} % remove notitlepage to move toc to a separate page than the title
\usepackage{graphicx}
\usepackage{subfigure}
\usepackage{amsmath}
\usepackage{braket}
\usepackage[colorlinks,linkcolor={blue},citecolor={blue},urlcolor={blue}]{hyperref}

\usepackage{array}
\usepackage{MnSymbol}

\usepackage{bbm} % this is for the command \mathbbm{}
\usepackage{amsfonts} % this is for the command \mathbb{}
\usepackage{chngcntr} % this is for \counterwithin command in the appendix
\usepackage[page,toc,titletoc,title]{appendix}

\usepackage{ulem} % this is for \sout{text}, i.e. strikethrough text

\usepackage{array}
\newcolumntype{P}[1]{>{\centering\arraybackslash}p{#1}}
\newcolumntype{M}[1]{>{\centering\arraybackslash}m{#1}}

\newcommand{\bB}{\mathbf{B}}
\newcommand{\br}{\mathbf{r}}

\newcommand{\bp}{\mathbf{p}}

\newcommand{\bv}{\mathbf{v}}

\newcommand{\bCalB}{\boldsymbol{\mathcal{B}}}
\newcommand{\bCalE}{\boldsymbol{\mathcal{E}}}

\newcommand{\CalB}{\mathcal{B}}
\newcommand{\CalE}{\mathcal{E}}

\newcommand{\bz}{\mathbf{z}}

\newcommand{\bD}{\mathbf{D}}

\begin{document}

\title{Ultrafast control of moir\'e pseudo-electromagnetic field in homobilayer semiconductors}
%\title{Dynamically controllable Moir\'e pseudo-electric field in homobilayer semiconductors}
%\title{Moir\'e pseudo-electric field in dynamically modulated homobilayer semiconductors}

\author{Dawei Zhai}

\author{Wang Yao}

\affiliation{Department of Physics, The University of Hong Kong, Hong Kong, China}
\affiliation{HKU-UCAS Joint Institute of Theoretical and Computational Physics at Hong Kong, China}

\date{\today}

\begin{abstract}
In long-wavelength moir\'e patterns of homobilayer semiconductors, the layer pseudospin of electrons is subject to a sizable Zeeman field that is spatially modulated from the interlayer coupling in moir\'e. By interference of this spatial modulation with a homogeneous but dynamically tunable component from out-of-plane electric field, we show that the spatial-temporal profile of the overall Zeeman field therefore features a topological texture that can be controlled in an ultrafast timescale by a terahertz field or an interlayer bias. Such dynamical modulation leads to the emergence of an in-plane electric field for low energy carriers, which is related to their real space Berry curvature -- the moir\'e magnetic field -- through the Faraday's law of induction. These emergent electromagnetic fields, having opposite signs at the time reversal pair of valleys, can be exploited to manipulate valley and spin in the moir\'e landscape under the control by a bias pulse or a terahertz irradiation. %Our study offers a simple scheme to realize emergent electrodynamics for valley/spintronics applications.
\end{abstract}

% insert suggested PACS numbers in braces on next line
%\pacs{}
\maketitle
\section{Introduction}
\begin{figure*}[ht]
	\includegraphics[width=6.5in]{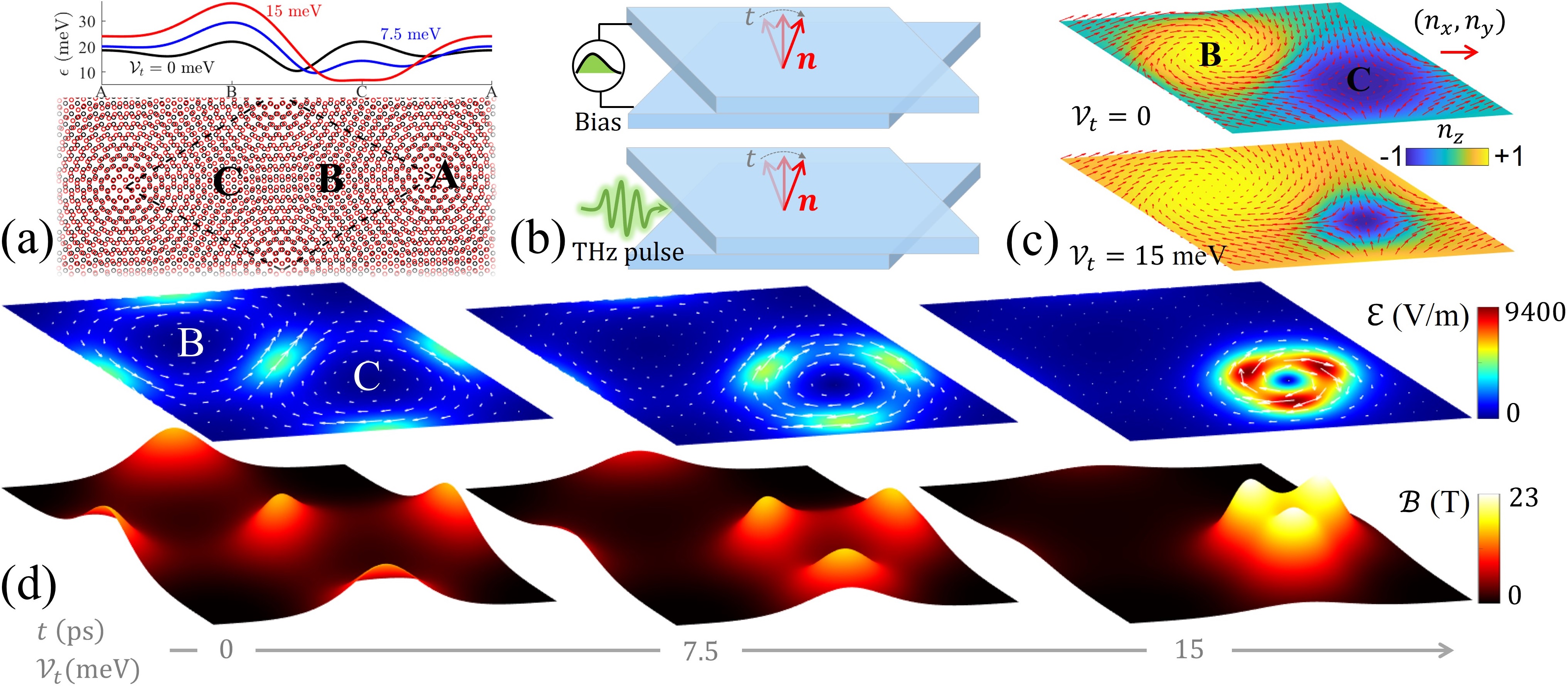}
	\caption{Emergent electromagnetic fields in a $0.5^\circ$ twisted moir\'e. (a) Lower: schematics of the moir\'e superlattice. A, B and C label the three high symmetry locales. A corresponds to the aligned parallel stacking; B (C) corresponds to the Bernal stacking with chalcogen (metal) atoms in the upper layer on top of metal (chalcogen) atoms in the lower layer. Upper: the moir\'e scalar potential $\epsilon=\mathcal{V}_0+|\vec{\mathcal{V}}(\br,t)|$ along the long diagonal of a supercell for $\mathcal{V}_t=0$ (black), $7.5$ meV (blue), and $15$ meV (red). (b) Schematics of a twisted bilayer with time-modulated interlayer bias (upper) or in THz field (lower), which introduces dynamics of $\vec{n}$. (c) Texture of $\vec{n}$ in a supercell, for $\mathcal{V}_t=0$ (upper) and $15$ meV (lower). In-plane (out-of-plane) component is represented by red arrows (background color). (d) Landscape of the emergent fields at three instants with $\mathcal{V}_t=0$, $7.5$, and $15$ meV. Electric field is in-plane as represented by white arrows, background color highlights its magnitude. Magnetic field is out-of-plane. $d\mathcal{V}_t/dt=1$ eV/ns.}
	\label{Fig:TwistFields}
\end{figure*}

Long wavelength periodic moir\'e patterns emerge by vertically stacking two monolayer crystals with a small orientation or lattice constant mismatch [Fig.~\ref{Fig:TwistFields}(a)].
The relative importance of kinetic energy and Coulomb interaction can be tuned in such superlattices, enabling various correlation-driven phenomena as pioneered by the discovery of superconductivity and correlated insulating states in magic angle twisted bilayer graphene~\cite{CaoYuanSuperconductivity,CaoYuanCorrelated}. 
Recently, moir\'e platforms composed of semiconducting transition metal dichalcogenides (TMDs) have also attracted great interests both for the moir\'e exciton optics~\cite{MoireExcitonHongyiSciAdv2017,MoireExcitonXiaodongXuNature2019,MoireExcitonXiaoqinLiNature2019,MoireExcitonFengWangNature2019,MoireExcitonFalkoNature2019} 
and the correlation phenomena~\cite{MottInsulatorTwistedWSe2WS2KinFaiNature2020,MottWignerTwistedWSe2WS2FengWangNature2020,MottInsulatorTwistedMoSe2ExcitonNature2020,WignerCrystalTwistedWSe2WS2KinFaiNature2020,MottInsulatorTwistedWSe2CoryDeanNatMater2020}.

Compared to graphene moir\'e superlattices, topology and correlation effects can be separately addressed in the moir\'e of TMDs. For example, in a heterobilayer, carriers are confined to a single layer by the band offset, whereas the moir\'e interlayer coupling simply manifests as a scalar superlattice potential, and the low energy moir\'e physics can be mapped to a Hubbard model tunable for exploring correlation-driven phases~\cite{WuFengchengHubbardPRL2018}. 
On the other hand, nontrivial topology can emerge in the moir\'e when the two constituent layers are strongly hybridized. This can be achieved in twisted homobilayer TMDs or heterobilayer with proper band alignment, where energy bands from the two layers are intertwined~\cite{WuMacDonaldPRL2019,HongyiPseudoFieldMoire,ZhaiPRMaterials,GeometricOriginTopologicalInsulatorTMDmoirePRB2021,TwistedHomobilayerWSe2FuLiang2021,TwistedMoTe2WSe2PNASFuLiang2021}. Topology and correlation can be brought together again in such cases. Recent experiments have shown signatures of electrically tuned phase transition from Mott insulator to quantum anomalous Hall insulator in such moir\'e superlattices~\cite{QAHTwistedMoTe2WSe2KinFai2021}.

The non-trivial topology in the strongly hybridized regime can also be understood from a real space picture, where the varying local atomic registries in moir\'e are imprinted onto the layer pseudospin degree of freedom of electron wave function. 
It has been shown that valence band edge carriers in near $0^\circ$ twisted homobilayer TMDs are described by $H=-\bp^2/(2m)+\vec{\sigma}\cdot\vec{\mathcal{V}}_m+\mathcal{V}_0$, where $\mathcal{V}_0$ is a small scalar potential for MoSe$_2$ considered in this work, $\vec{\mathcal{V}}_m$ is the moir\'e interlayer coupling and acts as a sizable Zeeman field on layer pseudospin $\vec{\sigma}$~\cite{HongyiPseudoFieldMoire,ZhaiPRMaterials,ZhaiPRL2020,WuMacDonaldPRL2019} (details in Sect.~S1 of Supplemental Material~\cite{SupplementalMaterial}). $\vec{\mathcal{V}}_m$'s magnitude and orientation $\vec{n}$ are both functions of the location $\br$ in the moir\'e, which exhibits vortex/antivortex patterns [Fig.~\ref{Fig:TwistFields}(c)]. 
For low energy carriers, their layer pseudospin will follow the local $\vec{n}$ in an adiabatic evolution, 
and consequently the center of mass (COM) motion is subject to an emergent pseudo-magnetic field (Berry curvature) 
\begin{equation}
\begin{aligned}
\bCalB_{\pm}
=\pm\frac{\hbar}{2e}\vec{n}\cdot(\partial_x\vec{n}\times\partial_y\vec{n})\,\hat{\bz}
\end{aligned},\label{Eq:B_in_terms_of_n}
\end{equation}
where $\pm$ is the valley/spin index. 
$\bCalB_\pm$ effectively realizes a fluxed superlattice, giving rise to the quantum spin Hall conductance of moir\'e mini-bands~\cite{WuMacDonaldPRL2019,HongyiPseudoFieldMoire,ZhaiPRMaterials}. Apart from the sign-dependence on valley/spin, it is analogous to the pseudo-magnetic field of an electron traversing a smooth magnetic texture, in which the electron spin adiabatically follow the local magnetization~\cite{NagaosaSkyrmionNatNano2013,NonAbelianGaugeDomainWallPhysRep2008}. Dynamics of magnetization also provide a context for its temporal control, but limited to a slow timescale.

In this work, we explore a unique possibility for ultrafast manipulation of the spatial texture of layer pseudospin underlying the nontrivial topology in strongly hybridized bilayer. The $z$ component of $\vec{\sigma}$ is associated with an electrical polarization, through which an out-of-plane electric field can sensitively introduce a homogeneous but dynamically modulated Zeeman term $\mathcal{V}_t(t) \sigma_z$. Its interference with the moir\'e interlayer coupling is exploited here to control the spatial-temporal profile of the overall Zeeman field $\vec{\mathcal{V}}(\br,t)=\vec{\mathcal{V}}_m(\br) + \mathcal{V}_t(t) \hat{\bz}$. Specifically, the ultrafast controllability of the Zeeman field orientation texture $\vec{n} (\br,t)$ by interlayer bias or terahertz (THz) pulse [Fig.~\ref{Fig:TwistFields}(b)] gives rise to a pronounced emergent in-plane electric field on electron's COM dynamics
\begin{equation}
\begin{aligned}
\CalE_{\pm,i}
=\pm\frac{\hbar}{2e}\vec{n}\cdot(\partial_i\vec{n}\times\partial_t\vec{n})
\end{aligned}\label{Eq:E_in_terms_of_n}
\end{equation}
with $i=x,y$, satisfying the Faraday's law of induction $\nabla\times\bCalE_\pm+\partial_t\bCalB_\pm=0$. 
We characterize such emergent electromagnetic fields and show how they drive carrier transport, and thus realize charge and valley/spin pump between different domains in the moir\'e, or control unidirectional valley/spin flow.

To conclude this section, we discuss the regimes for adiabatic approximation. There are two conditions as imposed by the smoothness of $\vec{n}$ in spatial and temporal space: (i) $\sqrt{E_kE_l}\ll2|\vec{\mathcal{V}}|$, and (ii) $E_t\ll2|\vec{\mathcal{V}}|$, where $E_k$ is kinetic energy of the electron, $E_l=\hbar^2/(2ml_c^2)$ and $E_t=\hbar/t_c$ denote the characteristic energy due to spatial and temporal variation with $l_c$ and $t_c$ representing respectively the characteristic length and time scale over which $\vec{n}$ varies (see Sect.~S5 of SM). In both conditions, $2|\vec{\mathcal{V}}|$ characterizes the local energy gap for the layer pseudospin to flip from $\vec{n}$ to $-\vec{n}$. In homobilyer TMDs, $|\vec{\mathcal{V}}|$ is typically in the range of 10--20 meV [see Fig.~\ref{Fig:TwistFields}(a) and discussions in the next section]. For $l_c\sim10$ nm, we have $E_l\sim0.6$ meV, thus condition (i) is usually satisfied for low energy carriers (e.g., $E_k$ as large as 40 meV corresponds to $\sqrt{E_kE_l}\sim5$ meV.). Condition (ii) can be respected for temporal modulation as fast as $\sim10$ THz.

\section{Rigid twisted moir\'e}
In the following, we will focus on $\bCalB_+$ and $\bCalE_+$ in $K$ valley, their subscripts will be neglected when no confusions arise.  
We consider the example of a $0.5^\circ$ twisted MoSe$_2$ bilayer. The black curve in Fig.~\ref{Fig:TwistFields}(a) shows the moir\'e scalar potential $\epsilon=\mathcal{V}_0 + |\vec{\mathcal{V}}(\br,t)|$ in the adiabatic motion when $\mathcal{V}_t=0$, while the other curves are the ones when homogeneous $\mathcal{V}_t$ is applied. It is worth noting that $\mathcal{V}_t$ can drive a topological phase transition, where the flux of $\bCalB$ per moir\'e supercell changes by a flux quantum~\cite{HongyiPseudoFieldMoire,ZhaiPRMaterials}. 
For the parameters adopted here, the transition occurs at a modest $|\mathcal{V}_t|=22.3$ meV (Fig.~S1 of SM~\cite{SupplementalMaterial}), in the neighborhood of which adiabatic approximation necessarily breaks down.  
This separates the two regimes where adiabatic evolution can be addressed. 
We will focus on the regime $|\mathcal{V}_t|<22.3$ meV in the main text, and results for $|\mathcal{V}_t|>22.3$ meV can be found in Sect.~S3 of SM~\cite{SupplementalMaterial}. 
The emergent electric field $\bCalE$ is controlled by the temporal profile of $\mathcal{V}_t(t)$. The numerical examples shown in the following are taken for $d\mathcal{V}_t/dt=1$ eV/ns, corresponding to gigahertz-frequency voltage control.

\begin{figure*}[ht]
	\includegraphics[width=7in]{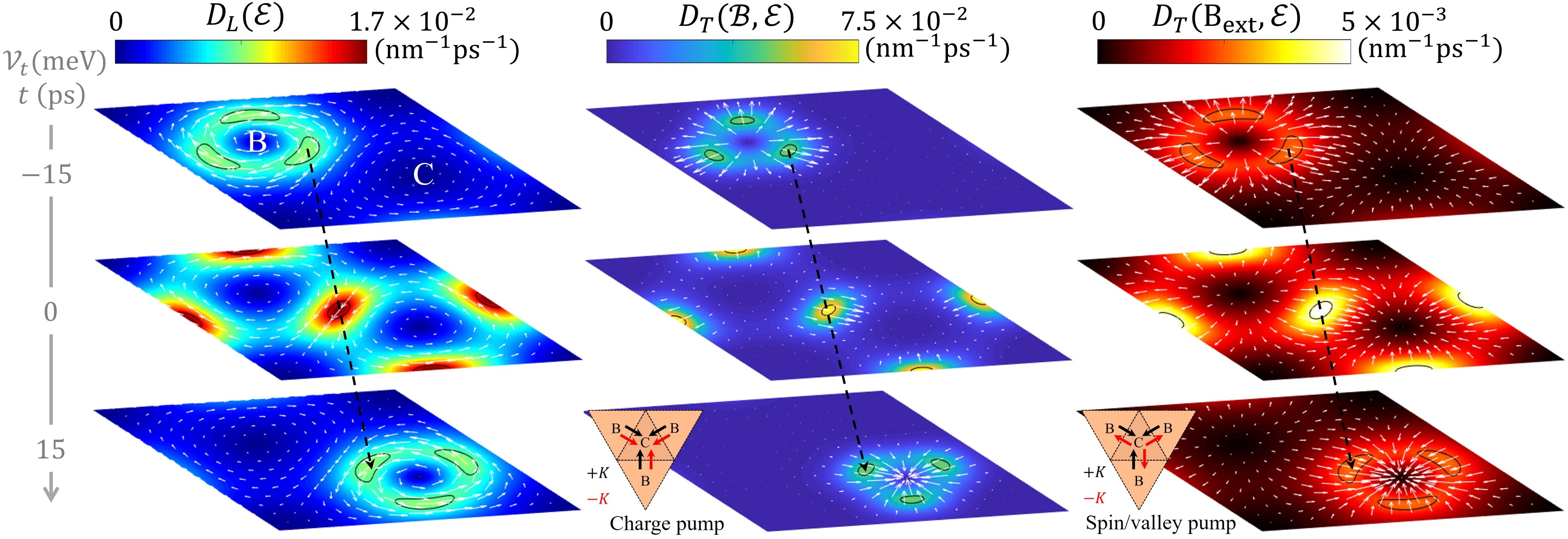}
	\caption{Local driving responses due to emergent electromagnetic fields in a $0.5^\circ$ twisted moir\'e. The three columns show distribution of $\bD_{L}$, $\bD_{T}(\bCalB,\bCalE)$, and $\bD_{T}(\bB_{\text{ext}},\bCalE)$ respectively in $K$ valley at different time. The white arrows represent the in-plane directions and background color denotes their magnitudes. Black ellipses are contours at 90\% of the peak values. The black dashed lines highlight the motion of driving hot spots with time. The black and red arrows in the insets illustrate parallel or counter flows in the two valleys, which result in charge or spin/valley current pumping respectively from B to C domains. Other parameters: $d\mathcal{V}_t/dt=1$ eV/ns, $\tau=1$ ps, $B_{\text{ext}}=1$ T.}
	\label{Fig:TwistCurrent}
\end{figure*}

Fig.~\ref{Fig:TwistFields}(d) show results of the emergent electromagnetic fields at three different instants with $\mathcal{V}_t=0$, $7.5$, and $15$ meV, respectively. We have set time axis such that $\mathcal{V}_t(t=0)=0$. The in-plane pseudo-electric field and the out-of-plane pseudo-magnetic field share a similar profile, as they are connected by Faraday's law. At $\mathcal{V}_t=0$, the fields are concentrated near the centers between B and C locales with a peak intensity $\sim4500$ V/m and $\sim15$ T, respectively. As $\mathcal{V}_t$ is increased, profile of the fields shows pronounced changes, where the hot spots move towards C locales. Intensities of the fields are also modified, e.g., the electric field increases by a factor of $\sim2$ at $\mathcal{V}_t=15$ meV. 

Instead of bias control, one can also apply THz pulses to modulate the Zeeman field texture [Fig.~\ref{Fig:TwistFields}(b) lower panel]. Intense THz transients can have a field strength of $0.1$ V/nm~\cite{THzFields}, which corresponds to $d\mathcal{V}_t/dt\sim100$ eV/ns, implying an $\bCalE$ that is two orders of magnitude stronger than those presented in Fig.~\ref{Fig:TwistFields}(d).
Strength of $\bCalE$ and $\bCalB$ also scale inversely with $L$ and $L^2$ respectively, $L$ being the moir\'e period.

The moir\'e pseudo-electromagnetic fields affect carriers in the same way as external electromagnetic fields, i.e., exerting Lorentz force on the COM motion. 
It is worth noting that Landau levels will not develop although the magnetic field has a fairly large peak intensity $\mathcal{O}(10)$ T. This can be understood by noticing that the field is nonuniform: the magnetic length $l_B=\sqrt{\hbar/e\CalB}$ is larger than the typical size of the spots where the field resides~\cite{ZhaiPRMaterials}.
For $\CalE\sim10^3$ V/m, a modest value on the field maps of Fig.~\ref{Fig:TwistFields}(d), the velocity of a carrier can be accelerated to $v\sim300$ m/s under the influence of the electric force within $1$ ps. This will induce a longitudinal ($\parallel \bCalE$) flow of carriers, which is a pure valley current, and also a spin current due to spin-valley locking~\cite{DiXiaoTMDPRL2012}.
And for a carrier with velocity $v\sim300$ m/s, the Lorentz force from a pseudo-magnetic field of $\CalB\sim10$ T has a comparable magnitude, which will deflect the motion towards the transversal direction ($\perp \bCalE$) with a Hall-like drift velocity $\propto\bCalE\times\bCalB$. Since both fields are valley-contrasted, a net charge current in the transversal direction is expected. 

From the above heuristic arguments, one expects that intense emergent electromagnetic fields will drive strong valley/spin and charge currents periodically distributed in the moir\'e superlattice.
However, quantitative evaluation of current responses become challenging here, as $\bCalB$ and $\bCalE$ are highly nonuniform. Furthermore, the spatially varying moir\'e potential landscapes [Fig.~\ref{Fig:TwistFields}(a)] also need to be taken into account in finding the instantaneous local current and carrier distribution under the dynamical control. 
In the following, instead of solving the instantaneous local current responses that need self-consistent consideration of carrier distribution, we focus on roles of the moir\'e pseudo-electromagnetic fields as driving forces for the motion of carriers based on semi-classical pictures.

First, we consider the valley-contrasted longitudinal drive, which can be estimated as
$\bD_{L}(\bCalE)=\rho \bv_L$, where $\rho$ denotes the local density of holes, $\bv_L=\tau e\,\bCalE/m$ is the velocity of a hole accelerated by $\bCalE$ in time $\tau$.
$\rho$ depends on the Fermi level $\epsilon_{F}$ and the location in the moir\'e potential landscape, i.e., $\rho (\br) \approx\frac{m}{2\pi\hbar^2}\left[\epsilon (\br)-\epsilon_{F}\right]$. 
For $\tau\sim\mathcal{O}(1)$ ps, $\bD_{L}(\bCalE)\sim\mathcal{O}(10)\times e \bCalE/h$ at $\epsilon-\epsilon_{F}\sim6.6$ meV.
Next, we consider the valley-independent transversal drive $\bD_{T}(\bCalB,\bCalE)=\rho\bv_T$. The transverse velocity due to the magnetic force can be estimated as $\bv_T=e\bv_L\times\bCalB\tau/m$, which yields $\bD_{T}(\bCalB,\bCalE)\approx\tau^2e^2\rho\,\bCalE\times\bCalB/m^2$.
Magnitudes of the two drives can be compared as $D_{T}(\bCalB,\bCalE)/D_{L}(\bCalE)=\tau e\CalB/m\sim3$ for $\tau\sim\mathcal{O}(1)$ ps and $\CalB\sim\mathcal{O}(10)$ T.
In the case of smoother spatial modulations and moderate electromagnetic fields, 
$\bD_{L}$ and $\bD_{T}$ correspond to the local current responses to the electromagnetic field in semi-classical transport theory~\cite{QianNiuRMP}. In addition to the intrinsic moir\'e pseudo-electromagnetic fields, one can also apply an external magnetic field $\bB_{\text{ext}}$ in the out-of-plane direction, which will result in a valley-contrasted transversal drive $\bD_{T}(\bB_{\text{ext}},\bCalE)$ as only $\bCalE$ reverses sign between the two valleys. 

Fig.~\ref{Fig:TwistCurrent} illustrates the instantaneous distribution of $\bD_{L}$, $\bD_{T}(\bCalB,\bCalE)$, and $\bD_{T}(\bB_{\text{ext}},\bCalE)$ at different times in $K$ valley as $\mathcal{V}_t$ is varied, with a hole density of $\sim0.019$ nm$^{-2}$ (averaged over the moir\'e supercell).
$\bD_{L}$ drives valley/spin currents with vortex patterns. 
$\bD_{T}(\bCalB,\bCalE)$ has the largest magnitude and drives charges flowing outward B locales towards C locales in the moir\'e. 
$\bD_{T}(\bB_{\text{ext}},\bCalE)$ due to a uniform out-of-plane external magnetic field (1~T) resembles $\bD_{T}(\bCalB,\bCalE)$, but exhibits opposite signs in the two valleys, so it pumps pure valley/spin flows from B to C locales. The hot spots of these current sources are also temporally controlled. For the example shown in Fig.~\ref{Fig:TwistCurrent}, the hot spots of $\bD$'s primarily occur near B locales at the beginning, forming a ring-like structure. The ring then splits into three petals, each of which moves towards the nearest C locale and merges into another ring, as sketched by the black dashed arrows.
\section{Strain engineered moir\'e and unidirectional valley flow}
In addition to twisting, lattice constant mismatch between the two layers can also yield moir\'e patterns.
Consequently, emergent electromagnetic fields and electron responses can also be obtained by applying heterostrain (i.e., layer dependent) to orientationally aligned homobilayers. Fig.~S5 of SM~\cite{SupplementalMaterial} shows similar moir\'e potential and emergent fields as those in Fig.~\ref{Fig:TwistFields} achieved in a moir\'e formed with biaxial strain. Large moir\'e supercells $\sim\mathcal{O}(100)$ nm -- desirable for the probe of the local currents and pumping effects -- can be achieved by applying a moderate strain ($\sim0.3\%$). 

\begin{figure}[t]
	\includegraphics[width=3in]{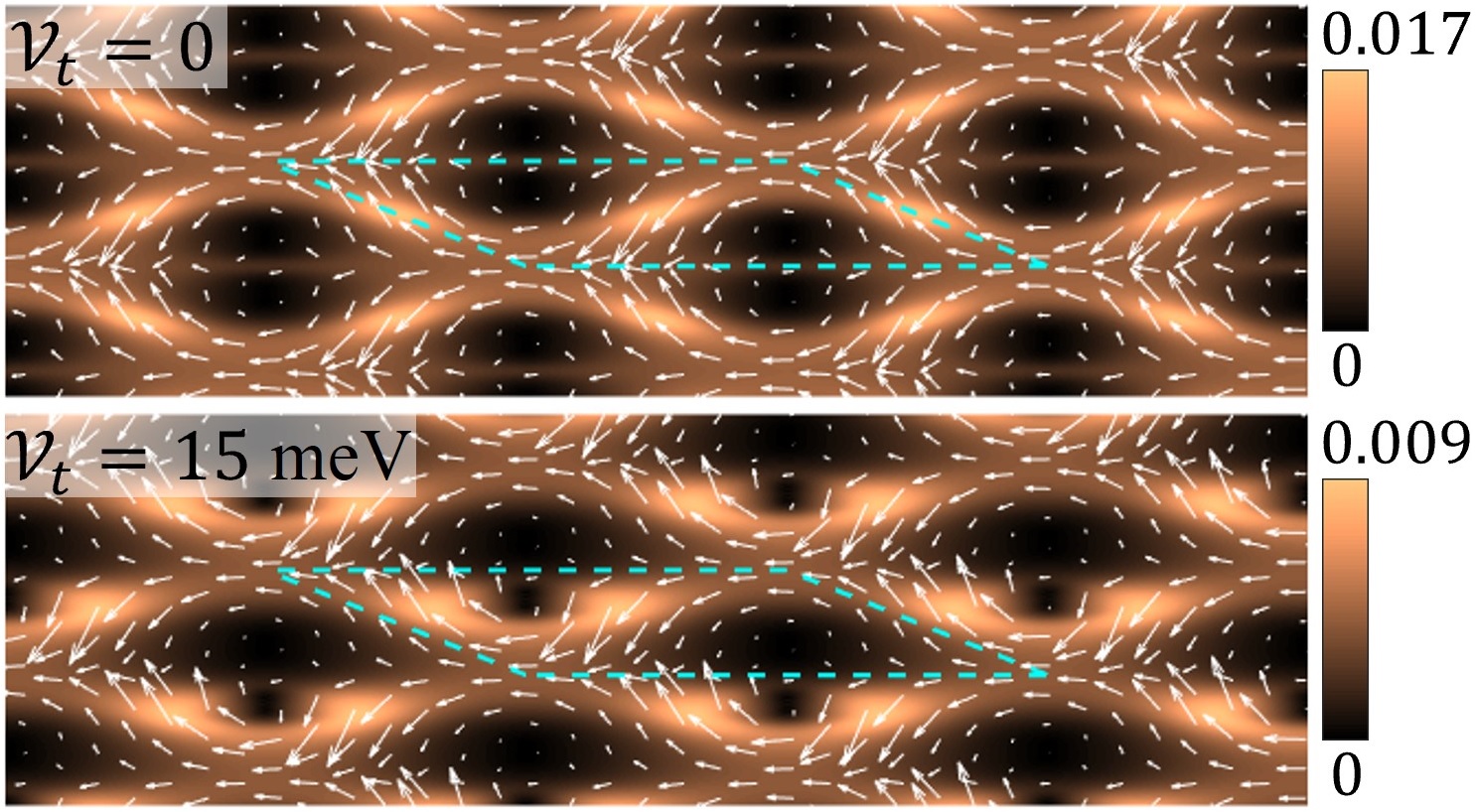}
	\caption{Unidirectional flow of pure valley/spin current driven by $\bD_{L}$ in a moir\'e from uniaxial strain. Arrows represent the in-plane directions and color map denotes the magnitude of $\bD_{L}$ (units: nm$^{-1}$ps$^{-1}$). Cyan dashed lines enclose one moir\'e supercell. Strain: 0.87\%, Poisson's ratio: 0.23. Other parameters are the same as those in Fig.~\ref{Fig:TwistCurrent}.}
	\label{Fig:Quasi1DCurrent}
\end{figure}

Moir\'e patterns with elongated geometries can also be engineered using uniaxial heterostrain. Fig.~\ref{Fig:Quasi1DCurrent} is an example of the moir\'e generated by a $0.87\%$ uniaxial strain, where the local $\bD_{L}$ responses are plotted.
In contrast to the circulating current patterns in the twisted moir\'e (Fig.~\ref{Fig:TwistCurrent}), the pseudo-electric field in this elongated moir\'e landscape drives a pure valley/spin current that has a net unidirectional flow determined by the strain direction. The fact that the left-right directionality of the driven valley/spin current is also controllable by the bias pulse shape (through the sign of $d\mathcal{V}_t/dt$) can be highly desirable for valleytronic applications.

\section{Relaxed twisted bilayer}
So far the discussions have been focused on rigid moir\'e lattices. Structure relaxation can be significant at small twisting angles, which will change the moir\'e landscape~\cite{RelaxationJainPRL2018,RelaxationKaxirasPRB2018,FalkoRelaxationTheory,FalkoRelaxationPRB2021,RelaxationNingWangNanoscale2021}.
We briefly discuss the effects of structural relaxation on the moir\'e pseudo-electromagnetic fields and transport in this section.

\begin{figure}[t]
	\includegraphics[width=3.4in]{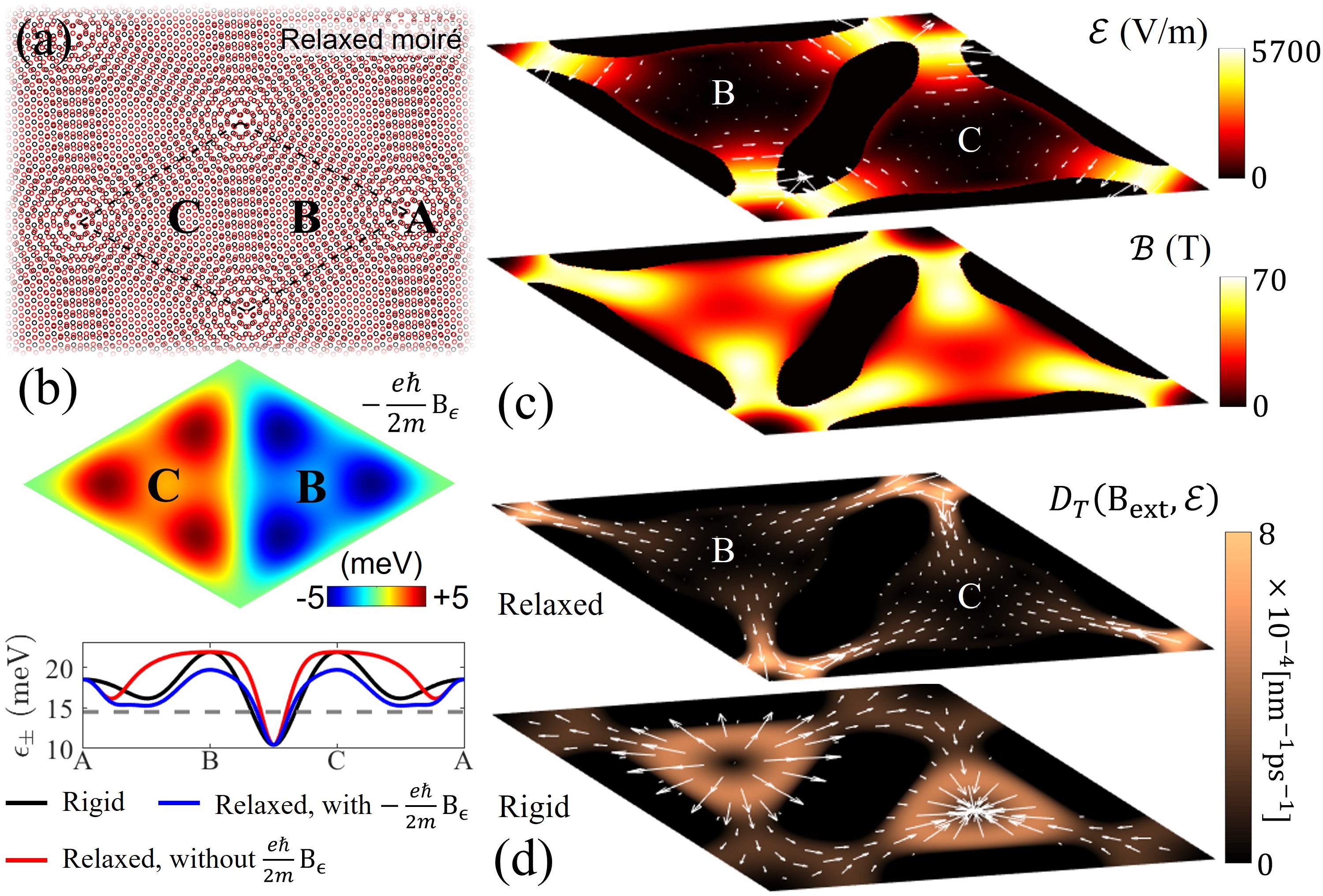}
	\caption{Emergent electromagnetic fields, and local driving responses in a relaxed twisted moir\'e. (a) Schematics of a relaxed twisted bilayer. (b) Strain Zeeman energy (upper panel), and modification of the moir\'e potential by variation of local atomic displacements and strain Zeeman energy (lower panel). The grey dashed line represents the Fermi level for upper panel in (d). (c) Distribution of the electromagnetic fields in a supercell. (d) Comparison of $\bD_{T}(\bB_{\text{ext}},\bCalE)$ for a relaxed (upper) and rigid (lower) moir\'e with $\mathcal{V}_t=0$. The white arrows represent the in-plane directions and background color denotes magnitudes. Other parameters: $d\mathcal{V}_t/dt=1$ eV/ns, $\tau=1$ ps, $B_{\text{ext}}=1$ T.}
	\label{Fig:RelaxedCurrent}
\end{figure}

As details of relaxation vary for different twist angles and materials, here we will focus on the qualitative features. In near $0^\circ$ twisted homobilayers, relaxation enlarges areas of B and C locales into large uniform triangular domains, while A locales will shrink. The resultant lattice shows narrow domain walls over which rapid spatial variation occurs [Fig.~\ref{Fig:RelaxedCurrent}(a)]. Such changes in the local stacking registries render two rather flat regions separated by sharp barriers in the moir\'e potential [Fig.~\ref{Fig:RelaxedCurrent}(b) red curves]. 
Atomic displacements due to relaxation introduce nonuniform strain into the moir\'e. In addition to changing atomic structures, the strain induces a pseudo-magnetic field $\bB_{\epsilon}$ with opposite sign at the two layers, which can contribute to the Lorentz force~\cite{ZhaiPRL2020}. 
It will also couple to the valley magnetic moment, introducing a Zeeman energy $-\frac{e\hbar}{2m}\text{B}_{\epsilon}$ [Fig.~\ref{Fig:RelaxedCurrent}(b)] and changes the layer pseudospin Zeeman field into $\vec{\mathcal{V}}(\br,t)=\vec{\mathcal{V}}_m(\br) + \mathcal{V}_t(t)\hat{\bz}-\frac{e\hbar}{2m}\text{B}_{\epsilon}(\br)\hat{\bz}$ (see Sects.~S6 and S7 of SM~\cite{SupplementalMaterial}). This further modifies the moir\'e potential as shown by the blue curve in Fig.~\ref{Fig:RelaxedCurrent}(b).

Adiabatic approximation breaks down near the domain walls due to the rapid spatial variations. So we focus on the emergent fields and the resultant carrier dynamics inside the flattened domains. Fig.~\ref{Fig:RelaxedCurrent}(c) illustrates the effects of relaxation on the emergent fields in the flattened domains taking $\mathcal{V}_t=0$ as an example. As the local atomic registries become uniform near B and C locales, $\bCalE$ can only exist near the boundaries of the cells, and its highest intensities reside near A locales. The pseudo-magnetic field as defined in Eq.~(\ref{Eq:B_in_terms_of_n}) shows similar distribution, but it is superimposed with the static strain induced contribution $\bB_{\epsilon}$ (Sect.~S6 of SM~\cite{SupplementalMaterial}). As a result the total pseudo-magnetic field $\bCalB$ becomes intensive and can reach several dozens of Tesla around B and C locales, where $\bCalE$ is minimal. 
In the following, we will look at $\bD_{T}(\bB_{\text{ext}},\bCalE)$ due to $\bCalE$ and a modest external magnetic field to illustrate the effects of structural relaxation on valley/spin pump. Results for $\bD_{T}(\bCalB,\bCalE)$ show qualitatively similar features.

Fig.~\ref{Fig:RelaxedCurrent}(d) upper panel shows results of $\bD_{T}(\bB_{\text{ext}},\bCalE)$ in $K$ valley in a relaxed moir\'e with $\mathcal{V}_t=0$ and hole doping with $\epsilon_F$ denoted by the grey dashed line in Fig.~\ref{Fig:RelaxedCurrent}(b), which corresponds to an average hole density of $\sim1.6\times10^{-3}$ nm$^{-2}$. The lower panel shows the results in a rigid moir\'e for comparison. Intensities of the driving become weaker due to moderate $\bCalE$ and low carrier densities. In a rigid moir\'e, the response forms a donut-like structure flowing outward (inward) around B (C) locales. As B and C locales flatten in the presence of structural relaxation, the hollow area of the current expands and the donut is split into three parts each moving towards the closest A locales. In this case, carriers are driven to move from B to C domains through the A locales as tunneling barriers.

\section{Discussions} 
The overall pumping effect by emergent electromagnetic fields depend on the shape of the interlayer bias/THz pulses. We elaborate this in the context of a temporal control by a bias pulse, which is composed of an upward slope followed by a downward slope [Fig.~\ref{Fig:TwistFields}(b) upper panel]. Although the bias returns to initial value after a pulse cycle, there could be a net accumulated pumping, which can be most intuitively seen in the ballistic regime as an example. Recall that the direction of $\bCalE$ is tied to the sign of $d\mathcal{V}_t/dt$ [Eq.~(\ref{Eq:E_in_terms_of_n})]. As a result, $\bCalE$ exhibits opposite signs when the bias ramps up and down, during which the electron is first accelerated and then decelerated. It should be noted that the electron keeps moving forward during the deceleration process, thus the pumping effect accumulated over a pulse cycle can be nonvanishing. When scattering is taken into account, the accumulated pumping per cycle shall depend on the shape of the bias pulse, while the pumping direction is determined by the sign of the pulse. These can be used for control of the net pumping effect. The situation becomes more complicated for THz pulses, where the field undergoes several oscillations in the pulse duration [Fig.~\ref{Fig:TwistFields}(b) lower panel]. The net pumping effect again depends on details of the pulse, and it is possible that the pumping direction can flip several times in a pulse duration. This may lead to an interesting scenario of generating AC pure valley/spin currents with quasi-1D directionality in the strained case, worthy of exploration for potential applications.

We would also like to compare with dynamical magnetic structures~\cite{NagaosaSkyrmionNatNano2013,NonAbelianGaugeDomainWallPhysRep2008} and materials subject to time-modulated deformations~\cite{DynamicalGrapheneVozmedianoPRB2013,DynamicalGrapheneFelixPRL2020,ChaoXingLiuReview2021}, where emergent electromagnetic fields can appear. Originated from the dynamical change of magnetization or the lattice structures, the controllability of pseudo-fields in these contexts is limited, and the modulation timescale is typically slow, where the Faraday's Law of induction suggests a small pseudo-electric field. In contrast, for carrier's layer pseudospin here, interlayer coupling in the moir\'e directly corresponds to a spatially modulated Zeeman field, while the applied out-of-plane electric field corresponds to a spatially homogeneous but ultrafastly tunable Zeeman field. The interference of the two components allows the Zeeman field texture to be spatial-temporally modulated without any structural change, in an ultrafast timescale by a THz field or an interlayer bias. The pronounced pseudo-electromagnetic fields and their controllability points to a new realm to explore emergent electrodynamics, as well as new opportunities to manipulate valley and spin in the moir\'e landscape.

\section{Acknowledgment}
We acknowledge fruitful discussions with Ci Li and Cong Xiao. This work is supported by the Research Grant Council of Hong Kong (17306819, AoE/P-701/20), and the Croucher Senior Research Fellowship.

\bibliography{TimeDependentRefs}
\bibliographystyle{apsrev4-1}

\end{document}